\documentclass[conference]{IEEEtran}
\IEEEoverridecommandlockouts
\usepackage{subfigure}
\usepackage{cite}
\usepackage{amsmath,amssymb,amsfonts}
\usepackage{algorithmic}
\usepackage{graphicx}
\usepackage{textcomp}
\usepackage{xcolor}
\usepackage{subcaption}
\def\BibTeX{{\rm B\kern-.05em{\sc i\kern-.025em b}\kern-.08em
    T\kern-.1667em\lower.7ex\hbox{E}\kern-.125emX}}
\begin{document}

\title{EMSpice 2.1: A Coupled EM and IR Drop Analysis Tool with Joule Heating and Thermal Map Integration for VLSI Reliability}
\author{
    \IEEEauthorblockN{Subed Lamichhane$^1$,  Haotian Lu$^1$ and Sheldon X.-D. Tan$^1$}
    \IEEEauthorblockA{$^1$Department of Electrical and Computer Engineering, University of California, Riverside, CA 92521}
    \IEEEauthorblockA{slami002@ucr.edu, hlu123@ucr.edu, stan@ece.ucr.edu}
    ~\thanks{
        \indent 
        This work is supported in part by NSF grant under No.CCF-2007135 and in part by NSF grant CCF-2305437. 
    }
}
\maketitle

\begin{abstract}
Electromigration (EM) remains a critical reliability concern in current and future copper-based VLSI circuits. 
As technology scales down, EM-induced IR drop becomes increasingly severe. While several EM-aware IR drop analysis tools have been proposed, few incorporate the real impact of temperature distribution on both EM and IR drop effects. 
In this work, we introduce {\it EMSpice 2.1}, an enhanced tool built upon the existing coupled IR-EM analysis framework, EMSpice 2.0, for EM-aware IR drop analysis. 
For the first time, {\it EMSpice 2.1} uniquely integrates Joule heating effects and practical thermal maps derived from actual chip conditions.
Additionally, it features improved interoperability with commercial EDA tools, facilitating more comprehensive EM and IR drop sign-off analysis. 
Our findings demonstrate that specific hotspot patterns significantly impact the lifetime of interconnects and overall chip reliability due to EM failures. 
Furthermore, our tool exhibits strong agreement with industry-standard tools such as COMSOL, achieving a speedup of over 200$\times$ while maintaining high accuracy.

\end{abstract}


\section{Introduction}
\label{sec:intro}

For nanometer VLSI chips with copper-based interconnects, increasing current density and shrinking interconnect dimensions, exacerbate reliability concerns related to electromigration (EM)-induced aging and failure. 
As technology advances, EM sign-off becomes essential steps to ensure the long-term reliability of integrated circuits. 
Therefore, the development of an accurate and efficient EM analysis tool is crucial for modern semiconductor design and verification.

The current EM models, such as those proposed by Black~\cite{Black:1969fc} and Blech~\cite{Blech:1976ko}, are facing criticism for being overly cautious and performing inadequately at the level of individual wire segments. 
Korhonen et al.\cite{Korhonen:1993bb} introduced a set of physics-based partial differential equations (PDEs) designed to compute the stress in each interconnect wire segment. 
Various analytical and numerical methods have been developed to solve Korhonen's equations for assessing 
EM-induced failures. Cook et al.~\cite{CookSun:TVSI'18} suggested a numerical approach,~\textit{FastEM}, based on the finite difference method (FDM). This method initially discretizes the PDE into linear time-invariant ordinary differential equations (ODEs) and then employs Krylov subspace-based reduction techniques to accelerate the computations. 
Furthermore, Chen et al.~\cite{ChenTan:TCAD'16} offered an analytical solution to the PDE, providing insights into the stress evolution within interconnect structures.


Most existing analytical and numerical solutions overlook the impact of spatial temperature gradients. 
Recent research highlighted in \cite{ChenTan:TCAD'21} suggests that thermomigration (TM) effects can rival those of electromigration. 
With advancements in technology leading to increased power density, greater Joule heating, and higher thermal resistances, these challenges are expected to intensify~\cite{KavousiChen:ASPDAC'22}. 
To mitigate these issues, Kavousi et al.~\cite{KavousiChen:ASPDAC'22} proposed an EM stress analysis method that incorporates the effects of thermomigration. 
However, this method assumes uniform boundary temperatures across all wires when calculating temperature distribution, which does not reflect the varied spatial temperatures found in actual chip operations.

In this paper, we introduce a full-chip TM-aware EM failure analysis tool designed for integrated physics-based EM-TM and electronic analysis. 
Our tool processes power grid netlists from the Synopsys ICC flow, delivering outcomes such as failed EM wires, their resistance changes, and the resultant IR drops in power grids over specified aging periods, considering Joule heating effects. 
A key innovation in \textit{EMSpice 2.1} is its ability to incorporate actual spatial temperature distributions across the chip into our evaluations. 

The novel contributions of this work include: 
\begin{itemize}
    \item Integration with standard EDA tools, allowing the use of standard post-layout parasitic extraction files to supply electrical and geometric layout information to the tool.
    \item Unlike other temperature-aware methods, our tool uniquely reads and utilizes the spatial temperature distribution of the chip, enabling more accurate temperature considerations for immortality filtering, Thermomigration assessments, and lifetime predictions.
    \item The tool offers a comprehensive array of User Interfaces (UIs)—including a Command Line Interface (CLI), a Graphical User Interface (GUI), and remote online portal—providing users with versatile options to input data and receive EM analysis results in interactive visual formats.
\end{itemize}




\section{Conceptual Review}
\label{sec:review}

EM occurs from the ongoing interactions between metal atoms and conducting electrons in power grids, leading to atom 
migration and resulting in tensile and compressive stresses at cathode and anode nodes. 
This can cause voids and hillocks that potentially lead to wire failures due to increased resistance or open circuits. 
Initially, the nucleation phase sees the formation of voids, followed by the post-voiding phase. 
Primarily, electromigration (EM), stress migration (SM), and thermomigration (TM) are key forces altering metal atom concentrations, heavily influenced by thermal effects~\cite{Abbasinasab:DAC'2018, KavousiTan:ICCAD'20}.

\subsection{Temperature modeling}

Starting with the first law of thermodynamics, as characterized by~\cite{carslaw_conduction_1959}, we can derive the stationary condition temperature distribution on a VLSI interconnect segment as~\cite{Lamichhane:ISVLSI'24}:
\begin{equation}
\small
    \begin{aligned}
        k_{cu}\nabla^2T - (Q_{jh}+Q_{conv}) = 0
    \end{aligned}
    \label{eq:stationary_thermal_equation}
\end{equation}
Here, $Q_{jh}=j^2\cdot\rho$ is the contribution from the Joule heating effect where $j$ is the current density in the interconnect and $\rho$ denotes resistivity of the metal wire.
$Q_{conv} =k_{cu}(T-T_0)/\Gamma^2 $ is the heat convection component, where $k_{cu}$ represents the thermal conductivity of metal wire, and $\Gamma$ is the thermal characteristic length influenced by the geometry and topology of the environment which can be estimated as described in~\cite{Abbasinasab:DAC'2018}:
\begin{equation}
\small
    \begin{aligned}
        \Gamma^2 = t_{cu}t_{ILD}\frac{k_{cu}}{k_{ILD}}
    \end{aligned}
\end{equation}
Where $t_{cu}$ refers to the thickness of wire, while $t_{ILD}$ and $k_{ILD}$ represent the thickness and conductivity of the dielectric.

Eq.~\eqref{eq:stationary_thermal_equation} can be resolved numerically through methods like FDM and FEM to obtain the spatial temperature distribution. Additionally, the temperature profile for a single wire, subject to boundary conditions $T(-L/2)=T_1$ and $T(+L/2)=T_2$, can be approximated as described in~\cite{Abbasinasab:DAC'2018, KavousiTan:ICCAD'20}:
\begin{equation}
    \small
    \begin{aligned}
       T(x) = T_0 + T_m \left[1-\frac{cosh\left(\frac{x}{\Gamma}\right)}{cosh\left(\frac{L}{2\Gamma}\right)}\right] + T_n \left[\frac{sinh\left(\frac{x}{\Gamma}\right)}{sinh\left(\frac{L}{2\Gamma}\right)}\right] 
    \end{aligned}
    \label{eq:temp_solution}
\end{equation}  
where $T_0=(T_1+T_2)/2,\ T_m=\rho\cdot j^2\cdot \Gamma ^2/k_{Cu}$ and $T_n=(T_1-T_2)/2$.
Existing methods use fixed values for $T_1$ and $T_2$ in temperature calculations, diverging from real-world conditions~\cite{KavousiChen:ASPDAC'22, Lamichhane:ISVLSI'24} that $T_1$ and $T_2$ may vary with time and are non-identical from wire to wire.
In contrast, \textit{EMSpice 2.1} derives $T_1$ and $T_2$ from actual temperature data, allowing for accurate $T(x)$ calculations for each interconnect segment using Eq.~\eqref{eq:temp_solution}.

\subsection{Temperature-aware immortal wires filtering}
During the nucleation phase of EM, if the steady-state stress at the cathode node ($\sigma_{\text{steady}}$) remains below 
the critical threshold ($\sigma_{\text{critical}}$), the interconnect tree is considered immortal. 
To assess immortality considering temperature effects, a voltage-based check is employed, as outlined by Sun and Demircan~\cite{SunDemircan:ICCAD'16} and Sun and Cook~\cite{SunCook:TCAD'18}. 
This work has been extended to consider the Joule heating in~\cite{KavousiChen:ASPDAC'22}, which will be used in this work. 
The only difference is that we will consider both Joule heating and given initial thermal distributions at the same time. 

\subsection{Coupled multiphysics analysis}
The electromigration (EM) stress, $\sigma$, accounting for electromigration (EM), stress migration (SM), and thermomigration (TM), is described by Korhonen's equation~\cite{DEORIO:Micro'2010}:
\begin{equation}
    \small
    \begin{aligned}
        \text{PDE: } \frac{\partial\sigma}{\partial t} & = \frac{\partial}{\partial x}\left[\kappa(x)\left(\frac{\partial\sigma}{\partial x}-S-M\right)\right], \quad t > 0
    \end{aligned}
\label{eq:tm_aware_korhonen_pde}
\end{equation}
\begin{equation}
\small
    \begin{aligned}
        \text{BC: } \kappa(x_b)\left(\left.\frac{\partial\sigma}{\partial x}\right|_{x=xb} - S-M\right) & = 0, \quad 0 < t < t_{\text{nuc}}
    \end{aligned}
\label{eq:tm_aware_korhonen_bc}
\end{equation}
\begin{equation}
\small
    \begin{aligned}
        \text{IC: } \sigma(x,0)  = \sigma_T
    \end{aligned}
\label{eq:tm_aware_korhonen_ic}
\end{equation}
Here, $S = \frac{eZ\rho j}{\Omega}$ denotes the EM flux, and $M = \frac{Q}{\Omega T}\frac{\partial T}{\partial x}$ represents the TM flux. The position-dependent diffusivity is described by $\kappa(x) =D_a(T(x))B\Omega/(k_bT(x))$, influenced by non-uniform temperature distributions. 
$D_a = D_0 \exp(-E_a k_b T(x)))$ is the atomic diffusion coefficient, with $D_0$ as a constant and $E_a$ as the activation energy. Additionally, $\sigma_T$ is the initial thermally induced residual stress, and $x_b$ marks the block terminals.

Korhonen's equation (Eq.~\eqref{eq:tm_aware_korhonen_pde}), incorporating electromigration (EM), stress migration (SM), and thermomigration (TM), is solved using the Finite 
Difference Time Domain (FDTD) method. This method analyzes stress over time in each multi-segment interconnect tree by transforming the differential equation, along with its boundary conditions (BCs) and initial conditions (ICs), into a linear time-invariant (LTI) system:
\begin{equation}
\small
    \begin{aligned}
        C\dot{\sigma}(t) & = A\sigma(t) + Bj(t) - D \\
        \sigma(0) & = [\sigma_1(0), \sigma_2(0), \ldots, \sigma_n(0)]
    \end{aligned}
\label{eq:em_tm_ode}
\end{equation}
In this model, $\sigma(t)$ is the stress vector, with $C$ and $A$ as $n \times n$ matrices. $D$ is an $n \times 1$ matrix for thermomigration (TM) effects, and $B$ is an $n \times p$ input matrix. 
A Krylov subspace method solves the ordinary differential equation (ODE) for stress evolution in a linear time-invariant (LTI) system, starting from initial stress $\sigma(0)$, which includes residual and previously computed stresses. 
The coupled EM-TM-IR drop equations addressed in this paper are structured as follows:
\begin{equation}
    \small
    \begin{aligned}
        &C{\dot{\sigma}(t)}  = A\sigma(t) + Bj(t) - D \\
        &V_{v}(t) = \int_{\Omega_L}^{}\frac{\sigma(t)}{B} dV \\
        &G(t)\times u(t) = Pj(t) \\
        &\sigma(0)  = [\sigma_1(0), \sigma_2(0), \ldots, \sigma_n(0)]
    \end{aligned}
\label{eq:em_tm_ir_ode}
\end{equation}
$V_v(t)$ estimates the void volume within a wire, $\Omega_L$ is the volume of the remaining wire, and $V$ is the wire's total volume. 
For IR drop calculations, a resistance network is derived from the power grid, and Modified Nodal Analysis (MNA) is used, detailed in Eq.~\eqref{eq:em_tm_ir_ode}. 
$G(t)$, the admittance matrix, changes over time with the wire resistance, influenced by electromigration (EM) and thermomigration (TM).

For the post-voiding phase, the PDE is identical to the nucleation phase, using the stress at $t=t_{nuc}$ as the IC. 
The BC at the void nucleation site is defined as follows:
\begin{equation}
\small
    \frac{\partial \sigma}{\partial x} \Bigg|_{x=x_{\text{nuc}}} = \frac{\sigma(x_{\text{nuc}}, t)} {\delta} , \quad t_{\text{nuc}} < t < \infty
\label{eq:lti_ode_nucleation}
\end{equation}
Here, $x_{nuc}$ indicates the void nucleation site, and $\delta$ is surface thickness. 
An LTI ODE system, similar to Eq.~\eqref{eq:em_tm_ir_ode}, is set up with this BC for the post-voiding phase.

During the post-voiding phase, as voids exceed the critical volume $V_{crit}$, wire resistance increases, affecting current density across the more resistive $T_a/T_aN$ barrier. 
This resistance change is approximated in~\cite{SunYu:TDMR'20},
\begin{equation}
\small
    \begin{aligned}
        \Delta R(t) = \frac{V_{\nu}-V_{crit}}{WH}\left[\frac{\rho_{T_a}}{h_{T_a}(2H+W)}-\frac{\rho_{C_u}}{HW}\right]
    \end{aligned}
\end{equation}
where $W$ denotes the wire width, $H$ and $h_{Ta}$ represents copper thickness and barrier layer thickness, respectively. $\rho_{Cu}$ and $\rho_{Ta}$ mark the resistivity of copper and $T_a/T_aN$ barrier, respectively.
During FDTD EM-TM simulations, $\Delta R(t)$ is recalculated at each timestep, along with current distribution, to reflect changes in resistance due to void growth.

\section{EMSpice 2.1 algorithm flow}
\label{sec:emspice2p0}

\begin{figure}[htb]
  \centering
  \includegraphics[width=\columnwidth]{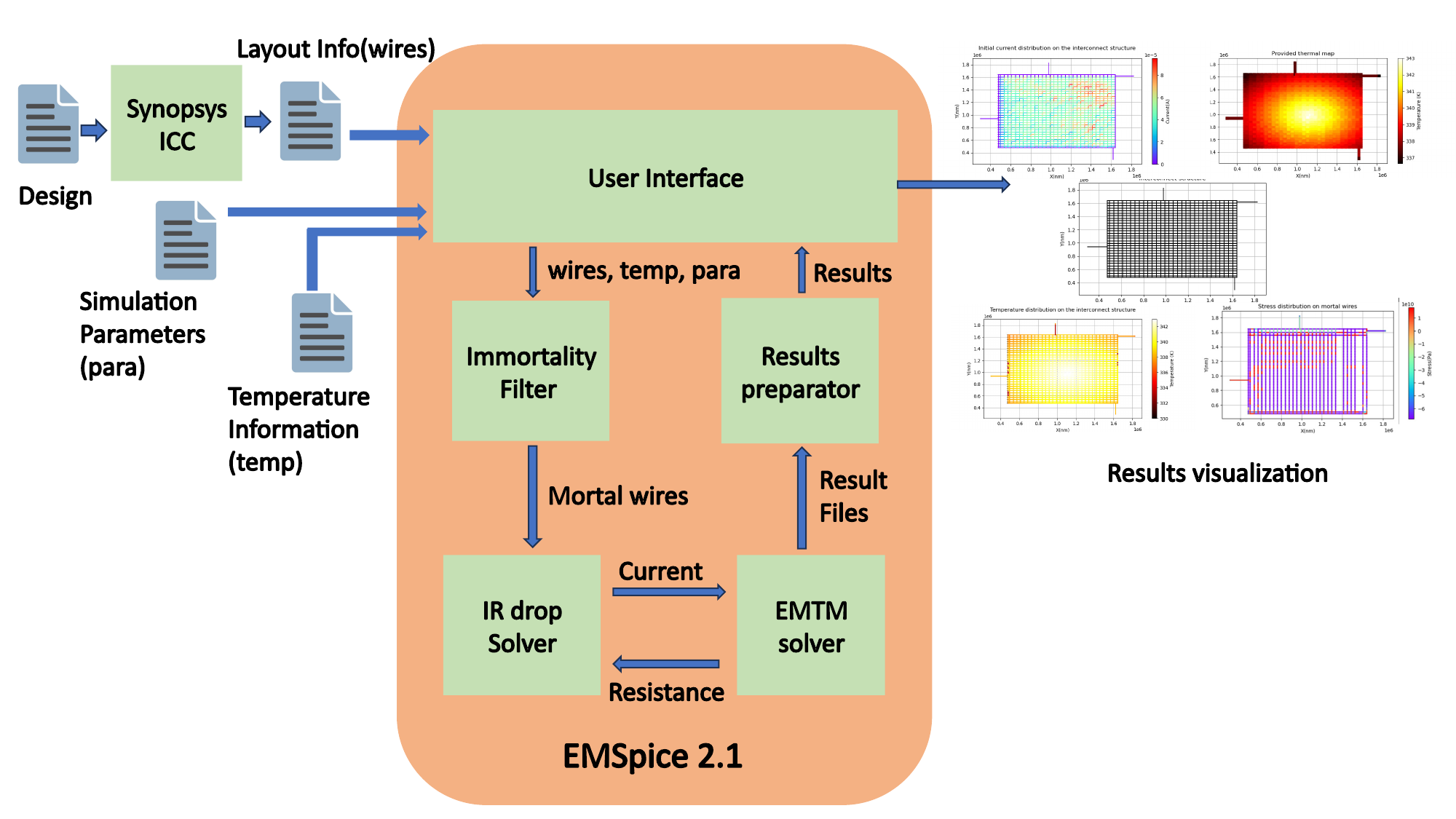}
  \caption{Overall framework of the proposed method }
  \label{fig:emspice2_framework}
\end{figure}

\textit{EMSpice 2.1} integrates with standard EDA tools like Synopsys and Cadence, as outlined in Fig.~\ref{fig:emspice2_framework}. 
The process begins with design layout and parasitic extraction using these tools. 
\textit{EMSpice 2.1} inputs the layout files and a parameters file detailing simulation time and material properties. 
Uniquely, it also accepts real temperature distribution files.

This tool offers three user interfaces (UI): a Command Line Interface (CLI), a Graphical User Interface (GUI), and an online portal for remote access without installation. 
Users input files through these UIs, which are then parsed to extract initial resistance, current information of the nets, initial temperature distribution, and simulation parameters, serving all other tool modules.

Techniques from Section~\ref{sec:review} filter out immortal trees that are impervious to electromigration effects. 
Mortal wires then proceed to the coupled FDTD-based EM-TM and linear network IR-drop solver, as detailed in Fig.~\ref*{fig:emspice2_framework}. 
This solver updates resistances based on time-dependent current densities affected by EM-TM effects. 
Simulation parameters like total time and step size are set in a user-provided parameters file. 
Post-simulation, the EM-TM solver outputs stress distributions and void locations in accessible formats. 
These results are processed for interactive visualization through the tool's UIs, as shown in Fig.~\ref{fig:emspice2_framework}.

\section{Numerical results and discussions}
\label{sec:observations}
\begin{figure}[h] 
    \centering  
    \vspace{-0.35cm} 
    \subfigtopskip=2pt 
    \subfigbottomskip=2pt 
    \subfigcapskip=-5pt
    \subfigure[The\ power\ grid\ structure\ of\newline Cortex-M0 design]{
	\label{fig:fig1}
	\includegraphics[width=0.5\linewidth]{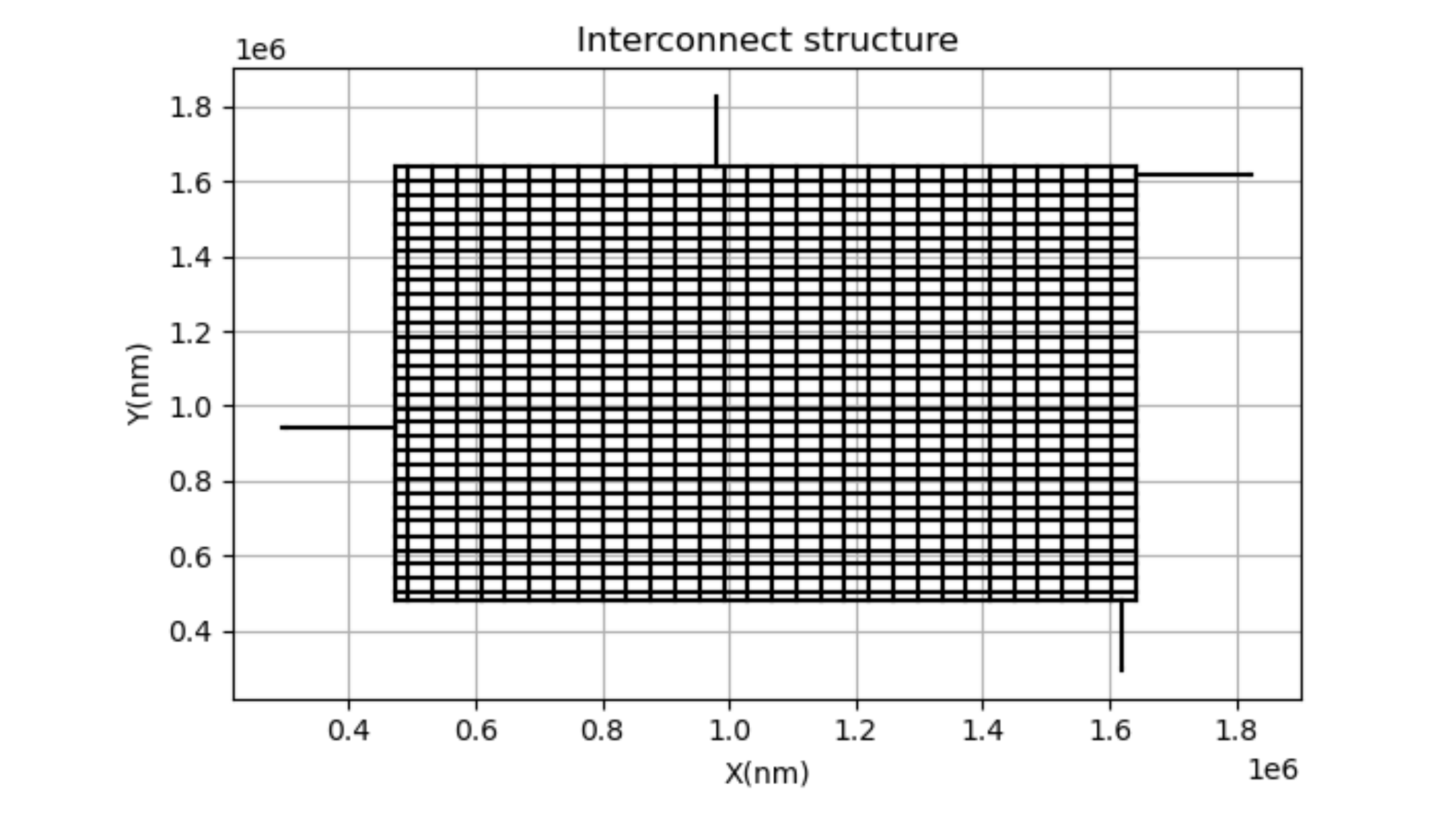}
    }
    \hspace{-0.9cm}
    \subfigure[The initial current density on the grid]{
	\label{fig:fig2}
	\includegraphics[width=0.52\linewidth]{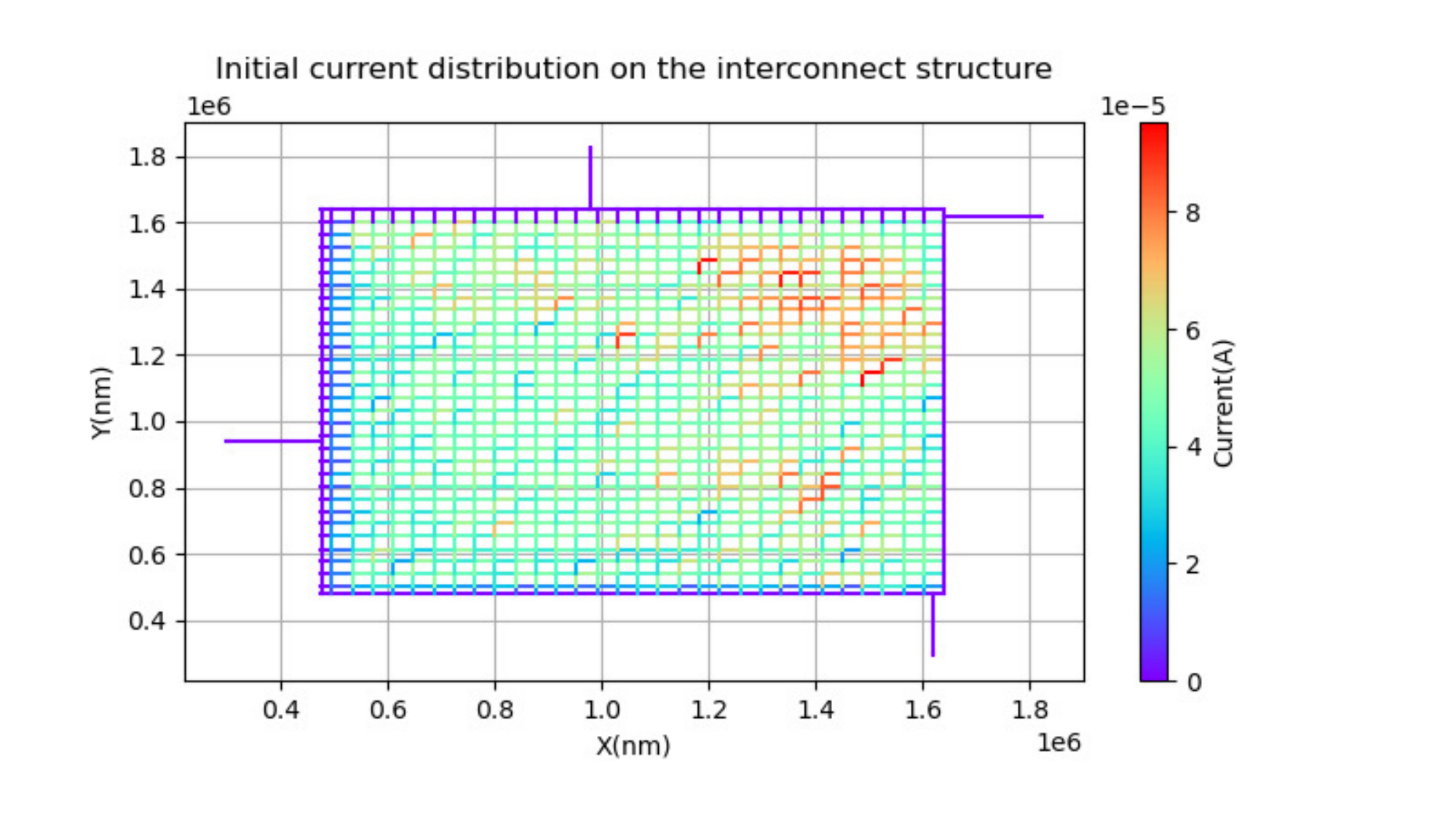}
    }

    \subfigure[The temperature distribution\newline with joule heating]{
	\label{fig:fig3}
	\includegraphics[width=0.52\linewidth]{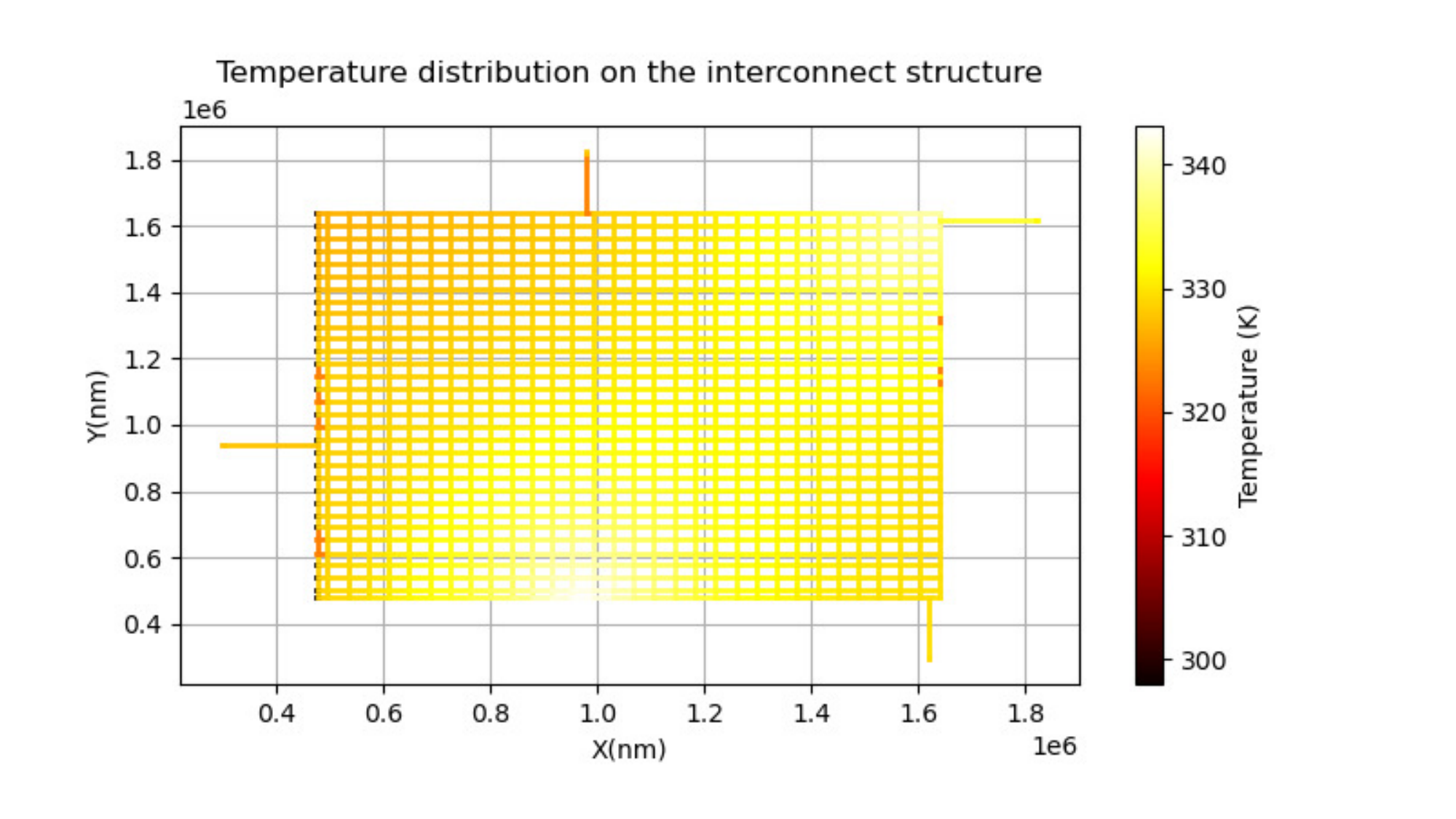}
    }
    \hspace{-1.12cm}
    \subfigure[The mortal wires and stress distribution on the wires for temperature distribution on the left, at $10^9$ seconds]{
	\label{fig:fig4}
	\includegraphics[width=0.52\linewidth]{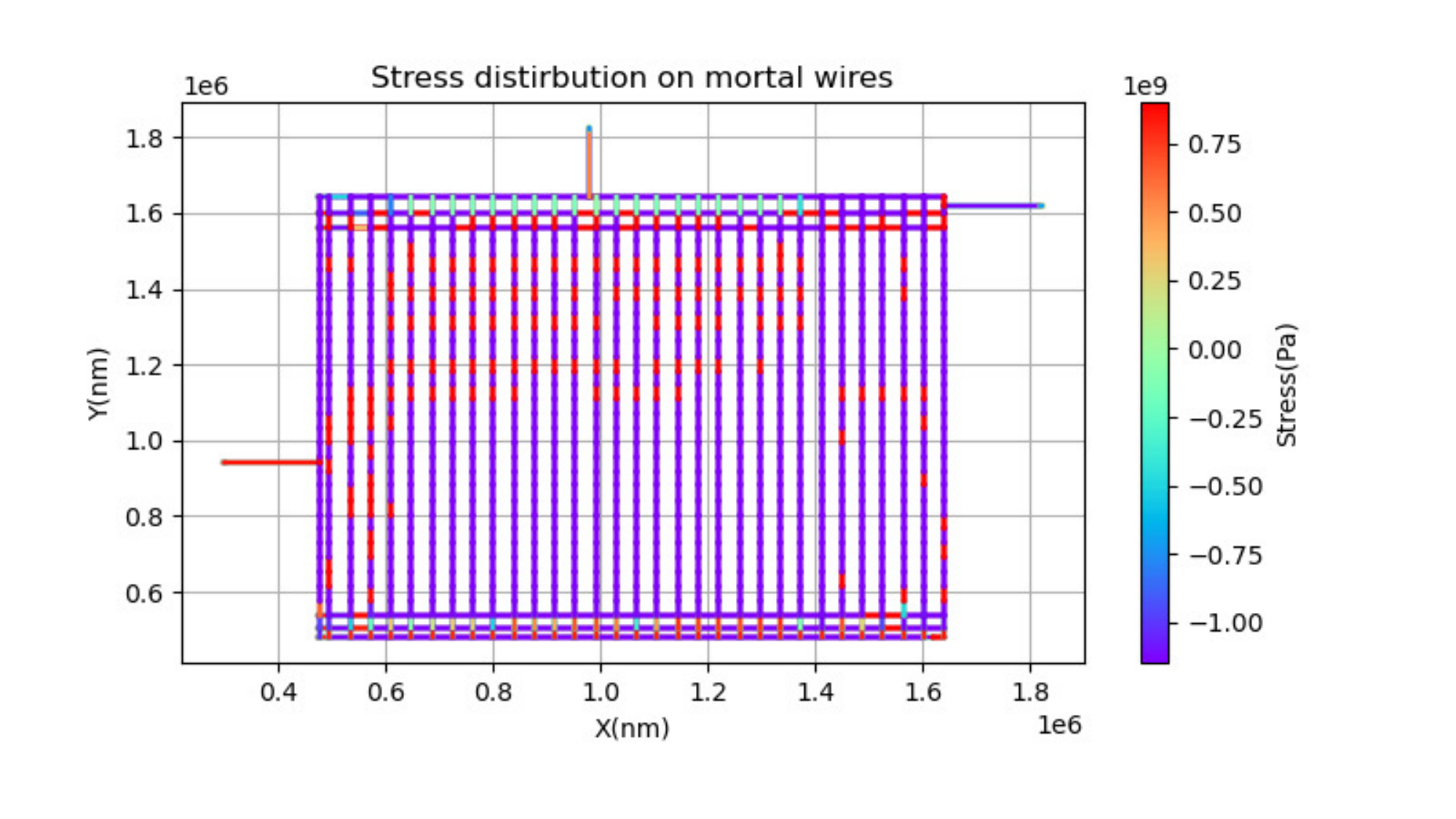}
    }

    \subfigure[The given temperature distri\newline-bution (case 1)]{
	\label{fig:fig5}
	\includegraphics[width=0.52\linewidth]{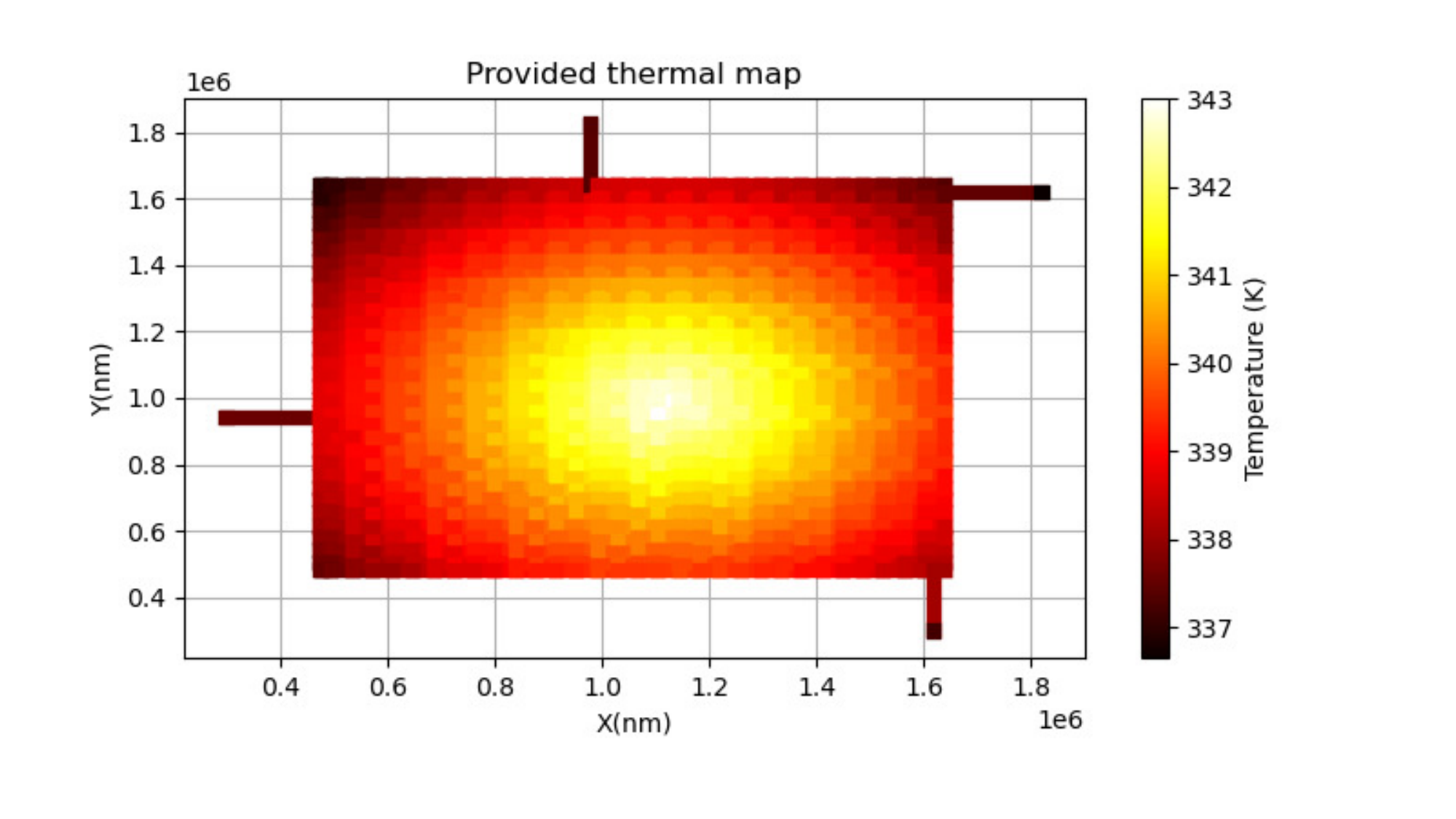}
    }
    \hspace{-1.12cm}
    \subfigure[The mortal wires and stress distribution on the wires for temperature distribution on the left, at $10^9$ seconds for case 1]{
	\label{fig:fig6}
	\includegraphics[width=0.52\linewidth]{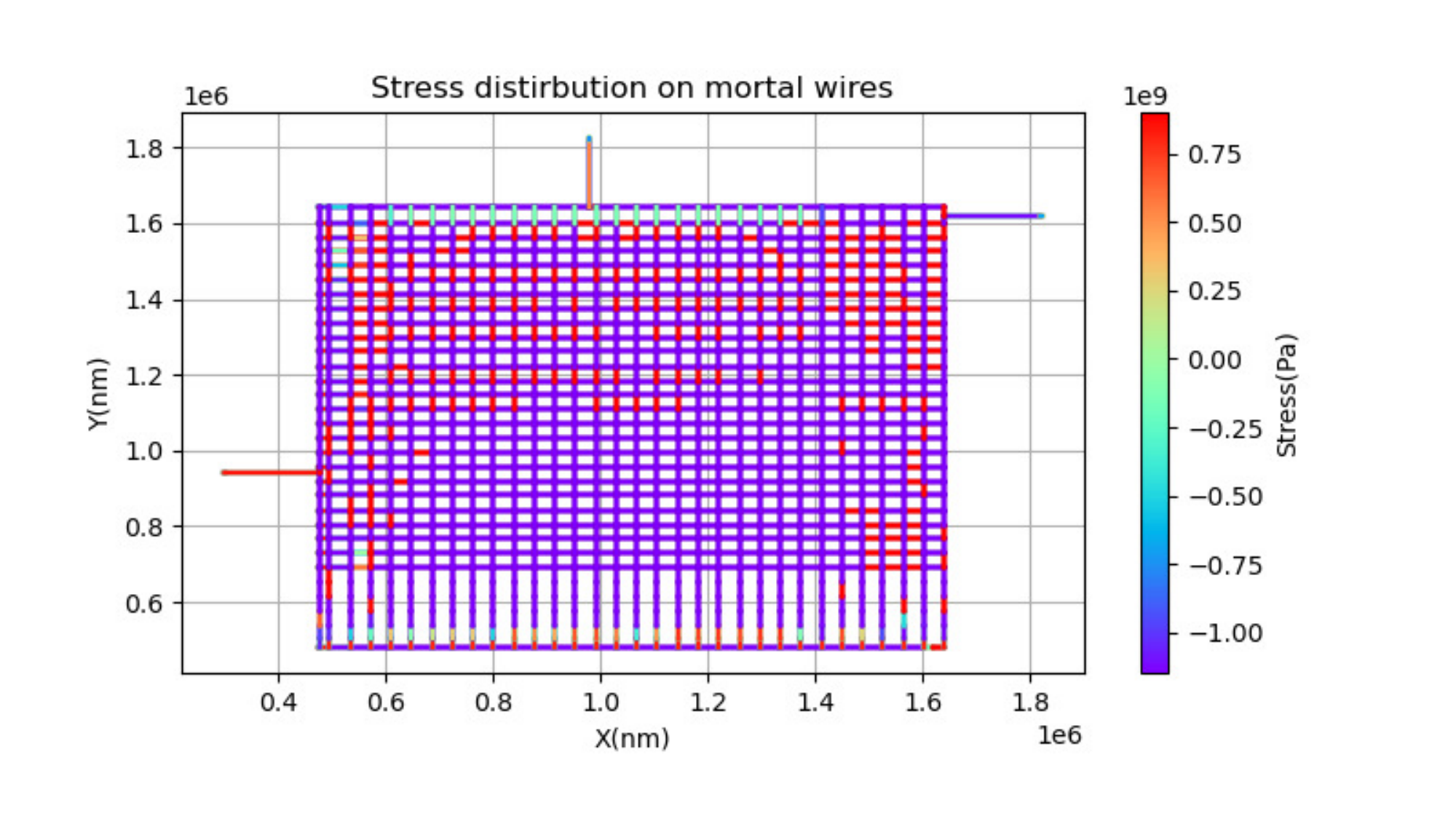}
    }

    \subfigure[The given temperature distri\newline-bution (case 2)]{
	\label{fig:fig7}
	\includegraphics[width=0.52\linewidth]{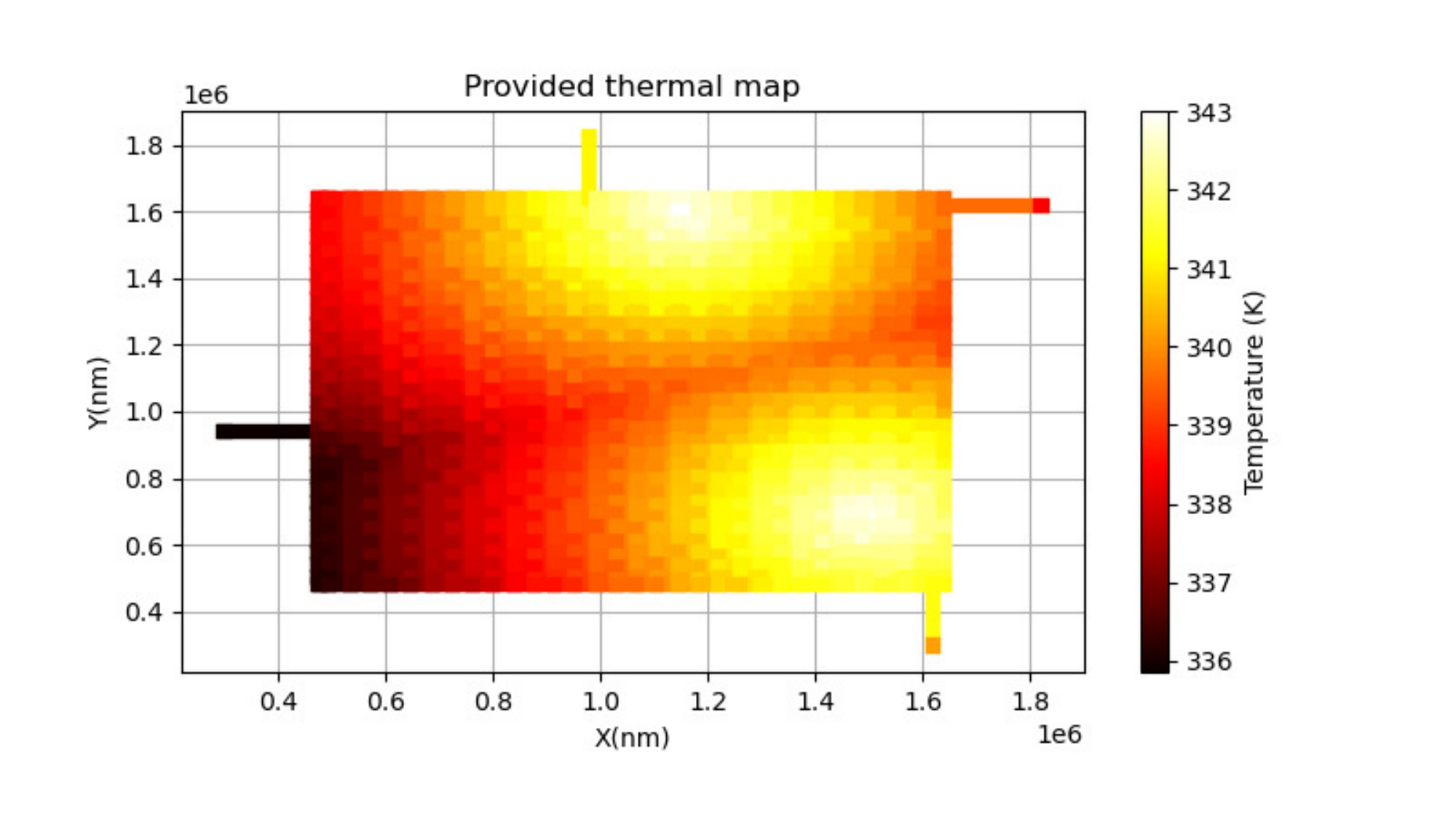}
    }
    \hspace{-1.12cm}
    \subfigure[The mortal wires and stress distribution on the wires for temperature distribution on the left, at $10^9$ seconds for case 2]{
	\label{fig:fig8}
	\includegraphics[width=0.52\linewidth]{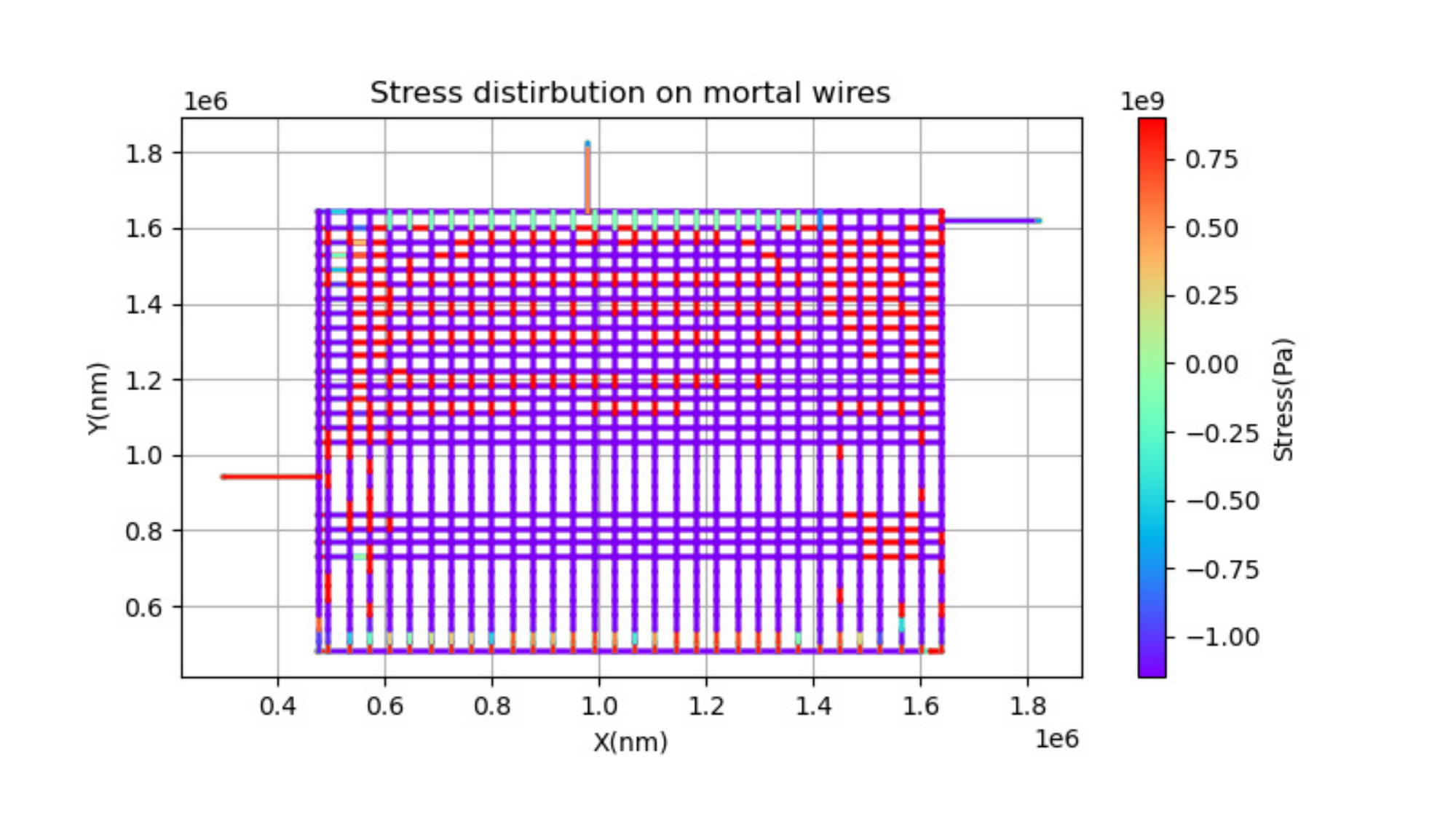}
    }
    \caption{The existing temperature map impacts on EM of power grid of Cortex-M0 design}
    \label{fig:all_figures}
\end{figure}

In this section, we present the numerical results from the proposed tool \textit{EMSpice 2.1} and examine temperature gradients' impact on interconnect mortality and lifetime of the chips. 
We use the \textit{Cortex-M0} power grid, depicted in Fig.~\ref{fig:all_figures}, as a case study.

\vspace{-5pt}

Fig.~\ref{fig:fig1} shows the design's power grid structure 
with 68 trees across two metal layers, averaging 31 segments 
per tree. Fig.~\ref{fig:fig2} displays the initial current 
distribution. Fig.~\ref{fig:fig3} depicts the temperature 
distribution, considering only internal Joule heating and an 
ambient temperature equal to the averages in 
Fig.~\ref{fig:fig5} and Fig.~\ref{fig:fig7}.

Fig.~\ref{fig:fig4} shows mortal wires and stress distribution at $1\times 10^9$ seconds, identifying 41 mortal wires out of 68. 
Temperature-aware filtering omits immortal wires, as discussed in Section~\ref{sec:review}. 
We evaluated power grid lifetimes under various temperature scenarios using $TTF= t_{life} = t_{nuc} + t_{inc} + t_{growth}$, 
following~\cite{SunYu:TDMR'20}. 
The lifetime was found to be approximately $5 \times 10^8$ seconds, or about 16 years. 

Fig.~\ref{fig:fig5} presents a thermal map case 1 with a distinct temperature gradient, leading to 58 mortal wires and a reduced lifetime of $3.8\times10^8$ seconds, or roughly 12 years. Fig.~\ref{fig:fig7} and Fig.~\ref{fig:fig8} show another temperature map case 2 and its impact on 51 mortal wires, with an average lifetime of about $4.4 \times 10^8$ seconds, or 14 years.

\begin{figure}[h]
    \centering 
    \vspace{-0.4cm} 
    \subfigtopskip=2pt 
    \subfigbottomskip=2pt 
    \subfigcapskip=-5pt
    \subfigure[Comparison of stress distribution result from the proposed method against COMSOL]{
	\label{fig:stress_dist}
	\includegraphics[width=0.8\linewidth]{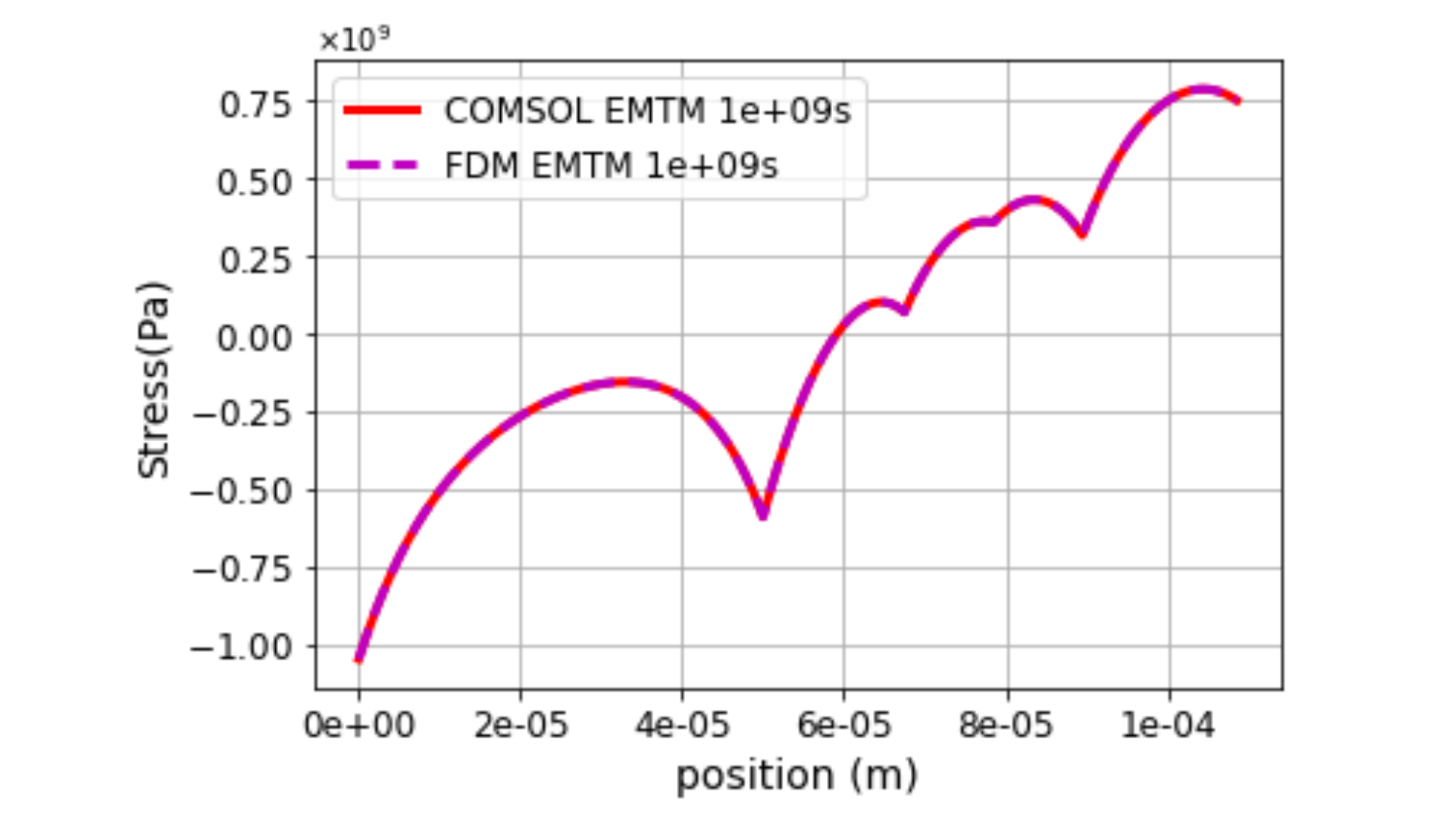}
    }

    \vspace{0.2cm}
    \subfigure[Temperature distribution used in the above stress calculation]{
	\label{fig:temp_dist}
	\includegraphics[width=0.8\linewidth]{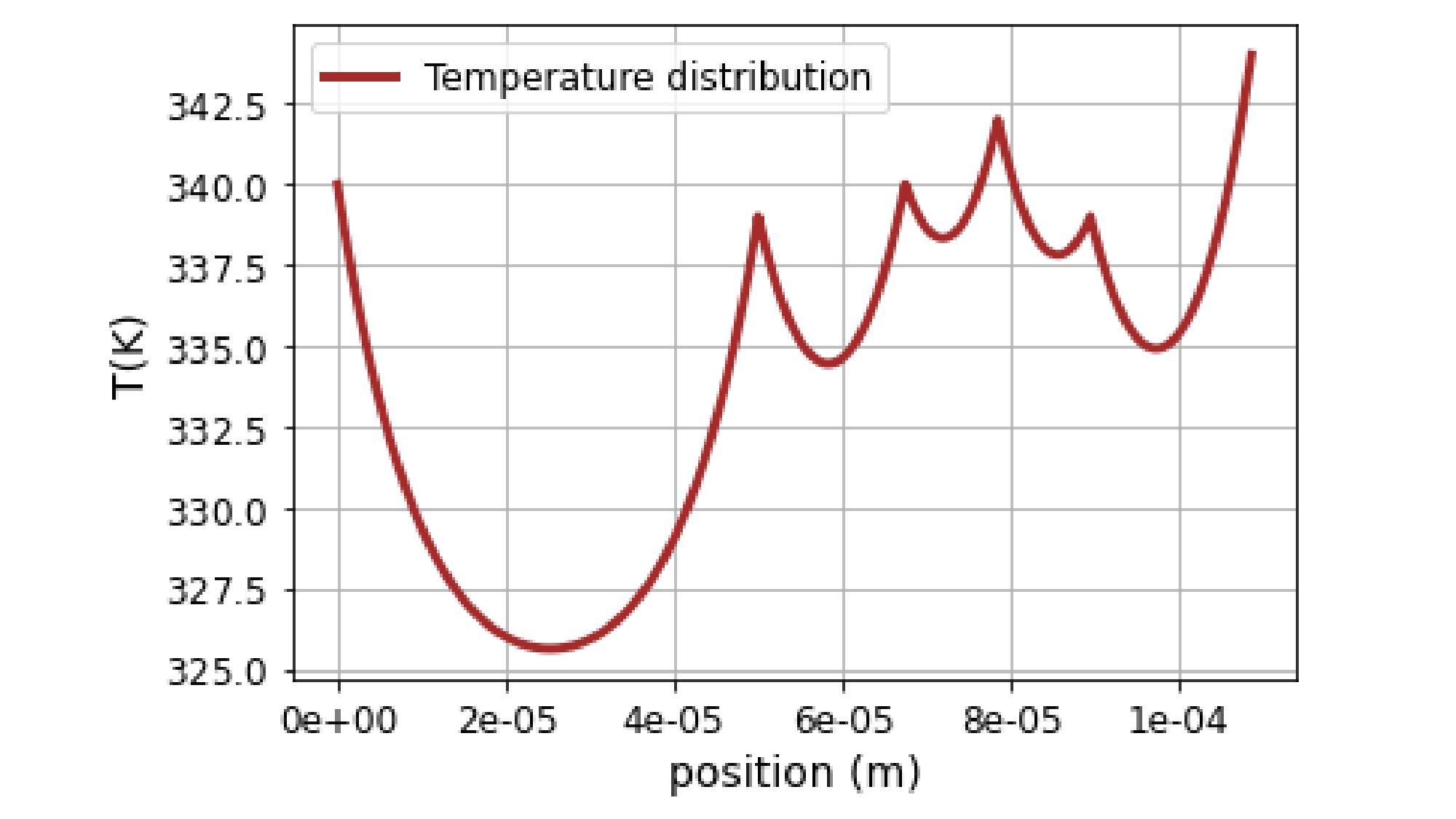}
    }
    \caption{Stress result comparison and temperature distribution used for the stress}
    \label{fig:stress_temp_dist}
\end{figure}

To assess the accuracy of our EM-TM stress solver, we analyzed 20 randomly selected trees from the \textit{Cortex-M0} power grid 
using COMSOL. 
The stress results from our temperature-aware FDM approach showed strong correlation with those from COMSOL~\cite{comsol:heat'2014}, as shown in Fig.~\ref{fig:stress_temp_dist}. 
We featured a five-segment tree from \textit{Cortex-M0} for clear illustration. 
Fig.~\ref*{fig:temp_dist} displays the temperature distribution within this tree. 
Stress levels ranged from $-1.1 \times 10^9$ Pa to $0.8 \times 10^9$ Pa. 
The RMSE for our method against COMSOL was $3.42 \times 10^7$, demonstrating an average error of only about 1.8\%.

\vspace{-5pt}

To further evaluate the performance of our method, we analyzed 20 randomly selected interconnect trees with 33 branches on average from the \textit{Cortex-M0} power grids using COMSOL. 
Our tool derived accurate solution in an averaged 1.6 seconds per tree, while COMSOL took about 315 seconds on the same task. 
This shows our efficient FDTD-based method achieves more than $200\times$ speedup over the FEM-based COMSOL.

\section{Conclusions}
\label{sec:conclusion}

In this paper, we introduced {\it EMSpice 2.1}, a coupled EM-aware IR drop analysis tool that, \textit{\textbf{for the first time}}, accounts for both electromigration (EM) and thermomigration (TM) by incorporating actual spatial temperature distributions. 
The tool uniquely integrates Joule heating effects and practical thermal maps derived from real chip conditions, enhancing accuracy in EM and IR drop analysis. 
Additionally, {\it EMSpice 2.1} offers improved interoperability with commercial EDA tools, enabling more comprehensive sign-off analysis. 
Our results show that specific hotspot patterns significantly affect interconnect lifetime and overall chip reliability due to EM failures. 
Moreover, our tool demonstrates strong agreement with industry-standard tools such as COMSOL while achieving over $200 \times$ speedup without compromising accuracy.

\bibliographystyle{ieeetr}
\bibliography{ref}

\end{document}